\documentstyle[12 pt,aasms4]{article}
\begin{document}
\title{Observations of a Correlated Gamma-Ray and Optical Flare for BL Lacertae}
\author{S. D. Bloom \altaffilmark{1}, D. L. Bertsch, R. C. Hartman, P. Sreekumar \altaffilmark{2}, and D. J. Thompson}
\affil{Laboratory for High Energy Astrophysics, NASA/Goddard Space
Flight Center, Greenbelt, MD 20771}
\author{T. J. Balonek, E. Beckerman \altaffilmark{3}, S. M. Davis, K. Whitman
\altaffilmark{4}}
\affil{Department of Physics and Astronomy, Colgate University, Hamilton, NY 13346-1398}
\author{H. R. Miller, D. Nair, L. C. Roberts, Jr.}
\affil{Department of Physics and Astronomy, Georgia State University, Atlanta, GA 30303}
\author{G. Tosti}
\affil{Cattedra di Astrofisica, Universit\'a di Perugia, Via A. Pascoli,
I-06100 Perugia, Italy}
\author{E. Massaro, R. Nesci, M. Maesano, F. Montagni}
\affil{Istituto Astronomico, Universit\'a ``La Sapienza'', via G. M. Lancisi,
29, I-00161, Rome, Italy}
\author{M. Jang}
\affil{Department of Astronomy and Space Science, Kyung Hee University,
Yongin, Kyungki-Do, 449-701, Korea}
\author{H. A. Bock, M. Dietrich, M. Herter, K. Otterbein, M. Pfeiffer, T. Seitz, S. Wagner}
\affil{Landessternwarte Heidelberg, K\"onigstuhl, 69117 Heidelberg, Germany}
\altaffiltext{1}{NAS/NRC Resident Research Associate}
\altaffiltext{2}{Universities Space Research Association}
\altaffiltext{3}{Visiting student from Wesleyan University}
\altaffiltext{4}{Visiting student from Cornell University}
\begin{abstract}
BL Lacertae was detected by the EGRET instrument on CGRO at the 10.2 $\sigma$ level with an average flux of $171  \pm 42 \times 10^{-8}\, {\rm photons \,
 cm^{-2} s^{-1}}$ at energies $> 100$ MeV
during the optical outburst of 1997 July. This flux is more than 4 times the previously highest level.
Within the July 15-22 observation
there was a dramatic factor of 2.5 increase
in the gamma-ray flux on July 18.75--19.08, apparently preceding by several hours
a brief optical flare. The gamma-ray flux decreased to its previous level
within 8 hrs and the optical flux decreased to its prior level in less
than 2 hrs.
The gamma-ray photon spectral index of 1.68 $\pm 0.12$ indicates that the spectrum during the 7-day observation was harder than the previous detection.
\end{abstract}
\keywords{Gamma Rays: observations --- galaxies: active --- BL Lacertae Objects: individual (BL Lacertae)}  
\section{Introduction}
BL Lacertae (z=0.069) is one of about 60 blazars that have been detected by the
EGRET instrument on  the {\it Compton Gamma-Ray Observatory} (CGRO) over the last 6 years (\cite{har97}). Though bright and variable across most of the
electromagnetic spectrum, it has not been an outstanding gamma-ray source
before this most recent outburst (\cite{cat97} and references therein).
This object was first detected at gamma-ray energies greater
than 100 MeV by EGRET in 1995 (\cite{cat97} and see Figure 1) but prior to 
this, BL Lac gamma-ray observations only yielded upper limits.
During a recent optical outburst of BL Lac (\cite{nob97}; \cite{mae97}) the CGRO was re-pointed to observe the BL Lac flare in gamma-rays.
BL Lac is the prototypcial object for a class of active galactic nuclei
(AGN) which typically have very weak emission and absorption lines 
and are characterized by variabilty in continuum emission and
polarization(\cite{pet97}). Approximately 20 \% of those blazars
detected by CGRO are BL Lac objects and the rest are flat spectrum
radio quasars (FSRQ's). Although some gamma-ray properties
are similar for these two classes, BL Lac objects tend to have lower
flux and less prominent variability than FSRQs (\cite{muk97}). 

The nature of the high energy gamma-ray emission of blazars has been disputed, with several models fitting data well for particular sources at specific
times (eg. \cite{har96}). However, most processes discussed in the literature are of the inverse Compton type. Inverse
Compton emission is a natural consequence of high soft photon densities in the vicinity of the relativistic electron
populations of blazar jets. If the seed photons are created via synchrotron
emission in the jet and then scatter off of the same population of electrons
which produced them (synchrotron self Compton or SSC; see \cite{blo96}), the gamma-rays will be observed to vary
simultaneously with the low energy photons (submillimeter through
optical). In the external scattering ``mirror'' model of \cite{ghi96} the 
optical outburst would lead the gamma-rays by roughly 1 day, since the optical emission is created via synchrotron emission within the jet and the 
optical photons are then scattered off of the broad line region clouds and
back into the jet, where they are scattered finally by the
relativistic electrons to produce high energy gamma-rays. The multiple 
relativistic Doppler boosts involved greatly enhance the observed gamma-ray emission. External scattering models which involve soft optical photons from 
an accretion disk (\cite{der93}) are much harder to test, since it is likely 
that the optical spectra of blazars are dominated by synchrotron emission 
from the jet, which ``washes out'' the optical emission from an accretion disk. However, in this case one would also expect optical emission leading the gamma-ray emission. 

Section 2 will summarize the gamma-ray and optical observations, and Section 3 
will discuss the results in light of the processes mentioned above.

\section{Observations}
During 1997 July 15-22, the EGRET instrument observed the region centered on BL Lac
in narrow field of view mode, which effectively looks at gamma-ray
sources within 20 degrees of the CGRO pointing axis. 
Details of the EGRET instrument can
be found in (\cite{har92}). Maps of the photon counts were created using the
standard techniques. We initially broke up the 7-day total exposure into
7  1-day maps, and then broken up the 4th 1-day map into 3 equal  8 hr
intervals once it was established that the gamma-ray emission peaked on the fourth day. 
A maximum likelihood analysis (\cite{mat96}) was performed
on the map to determine fluxes for individual sources, taking into account the galactic
diffuse background model of \cite{hun97} and the extragalactic diffuse model
of \cite{sre97a}. The gap of several hours seen between the points for
day 3 and day 4 of the gamma-ray observations (Figure 2) was caused by a 
scheduled loss of telemetry. 

Degradation of the EGRET spark chamber gas has led to
a decreased efficiency in detecting gamma-rays, mainly due to the reduced
capability of characterizing ``good'' gamma-ray events. Thus, 
scale factors have been applied to the current
data so that they are calibrated consistently with
data acquired at earlier points in the EGRET mission (see \cite{esp97} for
details of the in-flight calibration). 
These scale factors are determined by taking the ratio of the average intensity
of the diffuse emission from a well studied region of the sky (and at a given 
energy interval) from early in the mission and the diffuse emission from the
same region at the later time of interest.
Since the galactic diffuse emission is not expected to show any time 
variability, the diminished intensity is completely due to decreased instrument
performance. Thus, these scale factors represent an accurate relative
calibration of the EGRET data with time. Prior to this observation, a scale
factor of 1.6 was calculated.
We have now determined an additional scale factor
of 2.02 (total scaling is 1.6 $\times 2.02=3.2$) with an uncertainty of 20 \% determined by recent analysis. Of course, the amplitude of this
flare relative to previous measurements depends on the
size of this correction, but this does not affect relative
variations within a viewing period.

Visual magnitudes
were determined by several amateur observers, and these results were 
retrieved from
the American Association of Variable Star Observers (AAVSO) database (\cite{jmat97}) . Though all of the observers have used the same comparison stars,
many magnitude determinations were made by eye and thus there may be some
scatter in the measurements that is not intrinsic to the source
(these measurements also have large error bars). However,
in Figure 2, we see that there is agreement between visual magnitudes
determined by eye, and those determined by a CCD. R magnitudes measured 
using the 16 inch telescope of Foggy Bottom Observatory (Colgate), the
0.4 m telescope at Georgia State University, the 0.4m telescope at Perugia 
University Observatory, the 0.5 m telescope at Vallinfreda, Italy,
the 0.7 meter telescope at Landessternwarte Heidelberg and the 
Kyung Hee University telescope are
used along with
R magnitudes from \cite{mab97}. The $\vr$ color is approximately 0.7, and
we have applied  a $\Delta$m correction of 0.7 to the AAVSO data and our
V band data (from the Italian sites) to place
them on the same scale as the R band data.
We note that 
BL Lac was also detected by the OSSE instrument in the 50-300 keV range (\cite{gro97}) and was observed to
vary in the X-ray range at 2-10 keV with both RXTE and ASCA(\cite{mak97}; \cite{mad97}).
\section{Discussion}
From Figure 1 we can see clearly that the average gamma-ray flux is at least 
four times the previous value in 1995. Also, this measurement is twelve times
greater than the summed
95 \% confidence upper limit for phases 1-3 (\cite{cat97}).
Though the precise peak of the gamma-ray flare is uncertain, 
Figure 2 suggests that there is a lag of several hours between the peak
optical and peak gamma-ray measurements.
Unfortunately,
the sampling of the optical data does not completely rule out the possibility
of a very rapid optical flare, of larger amplitude than the observed optical peak, occuring during the peak of the gamma-ray flare
(or preceding it). However, the gamma-ray flux had already
decreased by the time of the peak optical magnitude.
We do note that since there are many other comparable
optical flaring events during this time period, we can not be certain
that the peak of the gamma-ray outburst is physically related to this
optical maximum. In addition, the exact value of the gamma-ray fluxes
are somehwat dependent on the arbitrary binning used here. 
Comparable effects have been seen for other blazars.
For the blazar 1406-076, the gamma-ray peak lagged the optical peak
by one day (\cite{wag95}). The 1991 June flare of 3C279 showed a similar
effect (\cite{har96}). However, none of these observations had 
optical data as densely sampled as this most recent observation of
BL Lac. In order to determine the best model for the high energy emission,
the relative lags must be determined accurately.

One possibility is that the gamma-rays are created via SSC emission from 
electrons accelerated in a front, possibly
a shock, moving through a jet (\cite{rom97}; \cite{mar96}) . However, if the optical emission lags the gamma-ray emission, then the optical flare can not be
the source of soft photons for the gamma-ray flare unless the
emitting region is initially optically thick to optical radiation.
Such extreme opacity conditions have not been previously observed for outbursts of blazars. The near simultaneity of the gamma-ray and optical peaks
(with the possibility of a delay of several hours) rules out external
scattering models mentioned above. We also note that the ratio of
the optical fluxes from this time period and that of January 1995 is
about 4.8, and that the ratio of the gamma-ray fluxes is 4.2. A simple
interpretation of the SSC process shows that equal enhancements of
the magnetic field and relativistic electron density would lead
to a greater amplitude of variation for the higher energy SSC flux, than
for the synchrotron flux, due to the extra factor of $N_O$ (electron
distribution normalization factor) in computing the SSC flux (\cite{blo96}).
However, if the Compton optical depth stays constant during the variations,
the amplitudes of variation for both synchrotron and SSC would be identical.

Figure 3 shows that the gamma-ray spectrum was harder during this flare,
as compared to that during the previous detection (a photon spectral index of 
$1.68 \pm 0.16$ for
the recent detection and $2.27 \pm 0.30$ for the earlier observation (\cite{cat97})). Spectral hardening during gamma-ray flares has been seen
previously for other blazars (\cite{sre97b}; \cite{muk97}) and is likely caused by injection of
higher energy  electrons which scatter soft photons preferentially to the
higher energies of the EGRET range. It is possible that second order
energy dependencies of the scale factor mentioned above could 
contribute, in part, to this spectral hardening.

To our knowledge, this is the first Target of Opportunity study 
of a blazar which has made extensive use of observations conducted by the
international amatuer astronomy community. We suggest that other investigators
make use of this valuable resource in future studies. We also
wish to stress the importance of including optical monitoring telescopes on
future X-ray and gamma-ray satellites, since it is likely that some of
the confusion over time correlation of flares would be resolved by
such instruments. 

We thank J. Mattox for many useful comments and for setting up a World Wide Web
home page supplying astronomers with up to date information on the BL Lac
outburst. 
HRM and ADN are supported in part by GSU's Research Promotion
and Enhancement Fund, and by grants from the Research Corporation,
and NASA (NAGW-4397).

\newpage
\figcaption[bllac_th.ps]{The complete time history of EGRET Gamma-ray Observations for BL Lacertae. Until Jan 1995, this source was not detected.
The arrows represent 95 \% confidence (2 $sigma$) upper-limits}
\figcaption[bllac.ps]{Optical and Gamma-Ray Light Curves for July 1997 Flare.
Both plots show that there was a peak on July 19. On the upper plot,
the open triangles represent visual magnitudes (recorded by eye)
corrected for $\vr$=0.7, asterisks are
V magntidues recorded by CCD with V filter, the open diamonds represent the
R magnitudes of Ma \& Barry, and the filled circles are the R and V band
data acquired by the authors at the sites mentioned in the text.
The dashed vertical line crosssing both
plots shows the end of the gamma-ray flare, for comparison to the optical
flare. }
\figcaption[bllac_410_6235spec.ps]{Gamma-Ray Spectra for BL Lac. The most
recent data (July 1997) suggests that the spectrum has hardened as compared to the
earlier measurement of (January 1995)}
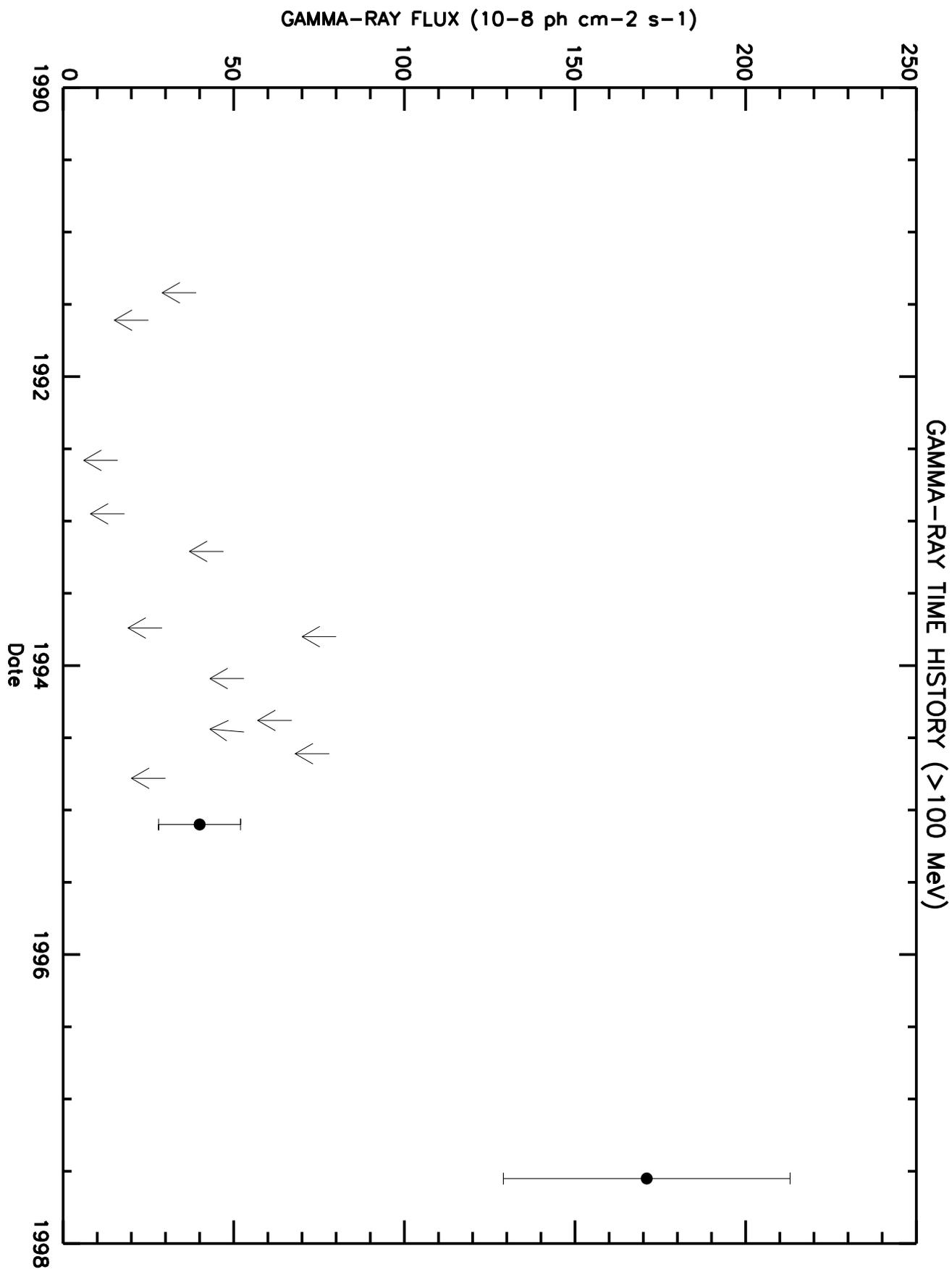
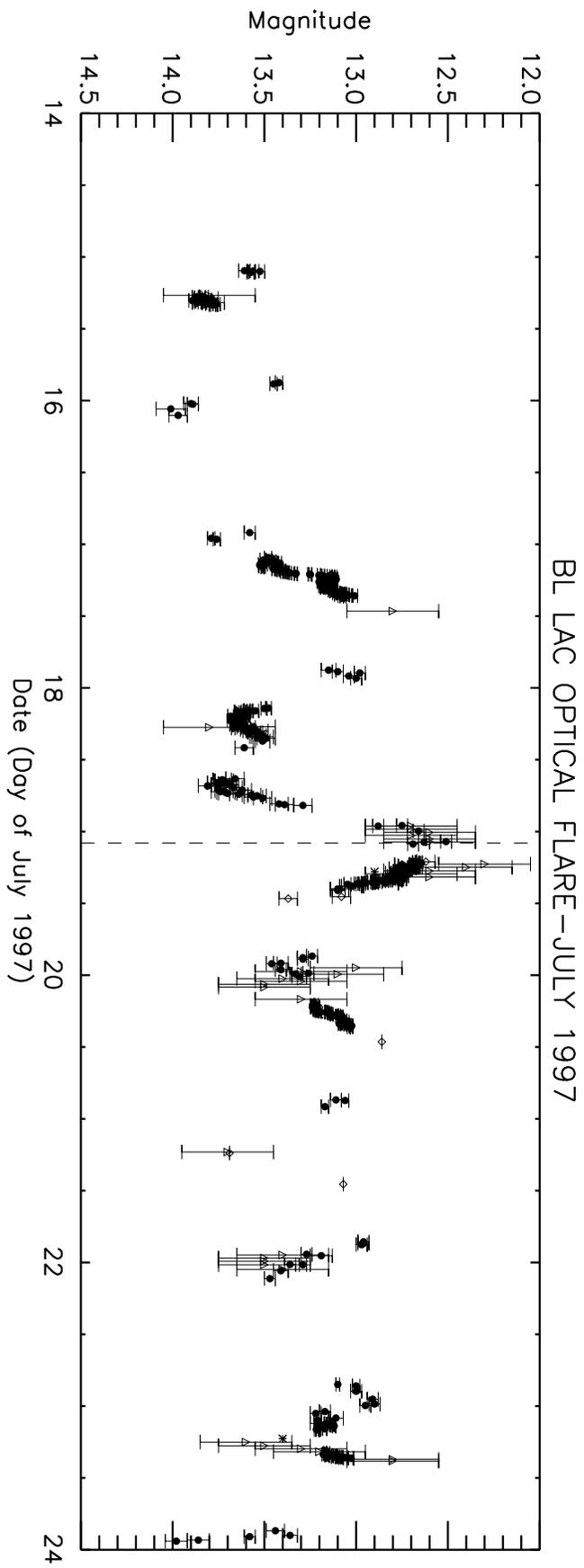
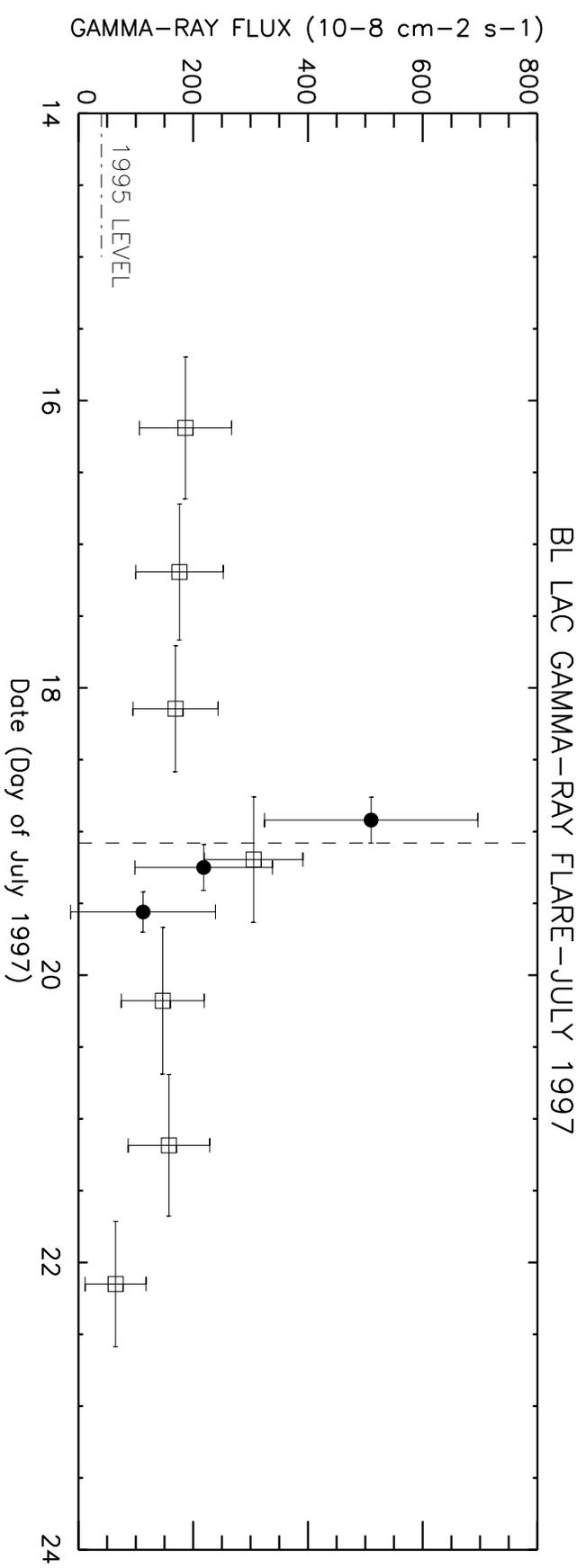
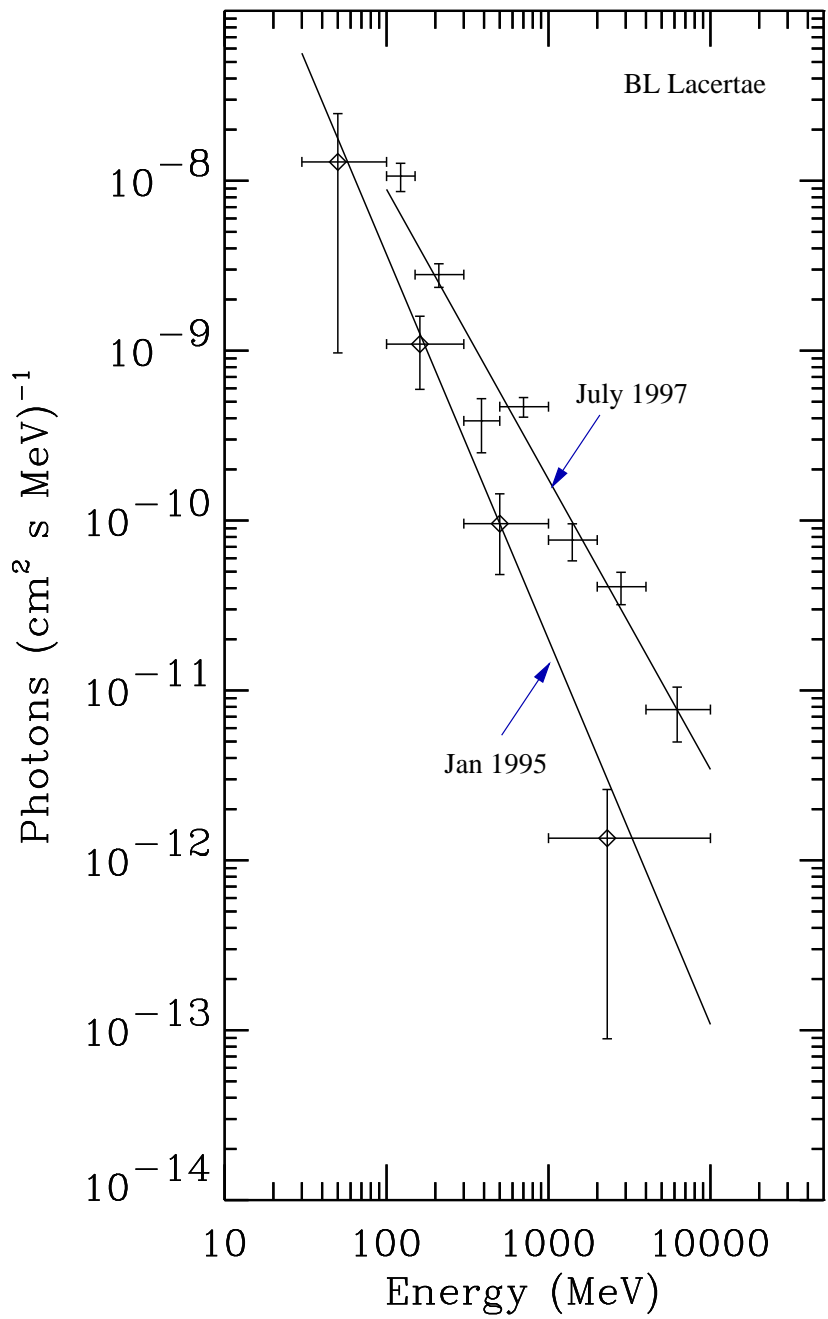
\end{document}